# Gravitomagnetic Barnett Effect

C.J. de Matos[*]

*ESA-ESTEC, Postbus 299, NL-2200 AG Noordwijk, The Netherlands*

M. Tajmar[†]

*Austrian Research Centers Seibersdorf, A-2444 Seibersdorf, Austria*

**Abstract**

Using the linearized theory of general relativity, the gravitomagnetic analogue of the Barnett effect is derived. Further theoretical and experimental investigation is recommended, due to the expected macroscopic values of the gravitomagnetic field involved in this effect, and to the constraints which would appear on quantum theories of gravity, currently under development, in case of non detection of the predicted phenomena.

---

[*] Staff Member, Science Management Communication Division, Phone: +31-71-565 3460, Fax: +31-71-565 4101, E-mail: clovis.de.matos@esa.int

[†] Research Scientist, Space Propulsion, Phone: +43-2254-780 3142, Fax: +43-2254-780 2813, E-mail: martin.tajmar@arcs.ac.at

# I - Linearized Theory of General Relativity

By linearizing Einstein's general relativity equations, gravitational analogues to the electric and magnetic fields are derived, respectively named the *gravitoelectric* $\vec{E}_g$, and the *gravitomagnetic* $\vec{B}_g$ fields[1]. Due to this similarity between gravitation and electromagnetism, it is possible to convert one field into the other one by applying certain conversion constants[2]. Now, if we have a certain mass distribution and flow, all that is necessary is to find a similar charge and current distribution in electromagnetic texts. We then use the formulas derived for the electric and magnetic fields and make the substitutions in the electromagnetic formulas to obtain the respective gravitational analogue.

Once we have calculated the fields generated by the mass density and currents, we can calculate the forces on a particle of mass $m$ by a force equation that is analogous to the Lorentz force equation,

$$\vec{F} = -m\nabla f - m\frac{\partial A_g}{\partial t} + m\vec{v} \times (\nabla \times A_g) \tag{1}$$

$$\vec{F} = m\vec{E}_g + m\vec{v} \times \vec{B}_g \tag{2}$$

If the test body is spinning and has an angular momentum $\vec{L}$, then the torque on it due to the gravitomagnetic field $\vec{B}_g$ will be by analogy (note that in the classical context the Landé factor $g = 1$)

$$\vec{N} = \frac{1}{2}\vec{L} \times \vec{B}_g \tag{3}$$

It should be emphasized that the previous discussion is approximate and is presented merely to provide a simple tool with which to make estimates and identify gravitational analogues of well known electromagnetic phenomena. By linearizing Einstein's equations the following assumptions have been made :

1. all motions are much slower than the speed of light to neglect special relativity,

2. the kinetic or potential energy of all bodies being considered is much smaller than their mass energy to neglect space curvature effects,
3. the gravitational fields are always weak enough so that superposition is valid,
4. the distance between objects is not so large that we have to take retardation into account.

We will show in the following, how this analogy can be used to calculate the gravitational analogue of the magnetic Barnett effect.

## II - Magnetic Barnett Effect

In 1915 S. J. Barnett[3] observed that a body of any substance (initially unmagnetized) set into rotation becomes the seat of a uniform intrinsic magnetic field parallel to the axis of rotation, and proportional to the angular velocity. If the substance is magnetic, magnetization results, otherwise not. This physical phenomenon is referred to as *magnetization by rotation* or as the *Barnett effect*.

Before we go further let us summarize some important basic concepts and definitions[4,5] related with the origin of magnetism in material media. The ratio between the angular momentum $\ell$ and the magnetic moment $m$ of the elementary magnetic gyrostat, $\ell/m$, is termed the *gyromagnetic ratio* and is denoted by $r$. For an electron describing a circular orbit $r = 2m/e$, for a spinning electron $r = m/e$, and for an electron system for which $g$ is the Landé splitting factor $r = \frac{1}{g}\frac{2m}{e}$. By means of gyromagnetic experiments, we measure $\frac{1}{g}\frac{2m}{e}$, and if we find that $g = 2$ we know that the elementary magnetic particle owes its properties to electron spin alone, whereas if we find that $g$ lies between 1 and 2 we know that orbital as well as electron spin motion is involved.

Let us suppose that the magnetic properties of a metallic (non-ferromagnetic) bar are due to electron systems each of angular momentum $\ell$, and that the bar is rotating with angular velocity $\Omega$ about its longitudinal axis $AB$. Each electron system which axis makes an angle $q$ with $AB$ will be acted upon by a couple $\ell\Omega\sin q$ tending to turn its axis of angular momentum parallel to $AB$, causing an increase in magnetization parallel to $AB$. Now, whether the gyrostats are turned by the application of an appropriate magnetic field $B$ parallel to $AB$ or by rotation of the bar, the result is the same, so that $Bm\sin q = -\ell\Omega\sin q$ or

$$B_{equi} = -\frac{1}{g}\frac{2m}{e}\Omega \qquad (4)$$

we may call $B_{equi}$ the equivalent magnetic field, the negative sign merely indicating reaction. The gyromagnetic experiment gives no information at all concerning the process of magnetization. Indeed, we might suppose that the impressed angular velocity $\Omega$ corresponds to the Larmor precession produced by the field $B$.

### III - Gravitomagnetic Barnett Effect

Using the isomorphism between gravitation and electromagnetism derived above from general relativity we are now able to compute the gravitational analogue of the magnetic Barnett effect. Therefore in Equ (4) we have to substitute $B$ and $e$ by $m$ and $B_g$ respectively and end with

$$B_{g\ equi} = -\frac{2}{g}\Omega \qquad (5)$$

Equ (5) tells us that any substance (initially ungravitomagnetized) set into rotation (see Fig 1) becomes the seat of a uniform intrinsic gravitomagnetic field parallel to the axis of rotation, and proportional to the angular velocity. If the substance is gravitomagnetic, gravitomagnetization results, otherwise not. This physical phenomenon shall be referred to as *gravitomagnetization by rotation* or as the *gravitomagnetic Barnett effect.*

The gravitomagnetic moment at the quantum level is defined as $\vec{m}_g = \frac{g}{2}\vec{\ell}$. Due to its definition, the Landé factor $g$ is a pure number (no dimension) and shall have the same value for gravitomagnetic and magnetic phenomena.

$$g = g_{GM} \tag{6}$$

Nevertheless, further research is needed to firmly establish this result. The only experiments using spin alignment to predict gravitational interactions are described in patents by H.W. Wallace[6,7]. His macroscopic observations, supporting our simple derivation, however have never been reproduced or seriously investigated in the literature up to now.

**IV Discussion**

In nature, every electromagnetic field is associated with a gravitic-gravitomagnetic field depending on the charge-to-mass ratio from the source particle[2]

$$\vec{B}_g = -\frac{m_{0g}}{m_0}\frac{m}{e}\vec{B} \tag{7}$$

$$\vec{g} = -\frac{m_{0g}}{m_0}\frac{m}{e}\vec{E} \tag{8}$$

However, due to the possibility of having neutral matter, only the gravitomagnetic field can exist without an associated electromagnetic field. Magnetization and gravitomagnetization appear also generally together[2]

$$\vec{M}_g = \frac{m}{e}\vec{M} \tag{9}$$

If magnetization and gravitomagnetization results from the application of a magnetic field to the body the relation between the magnetic and gravitomagnetic susceptibility is:

$$c = c_g \tag{10}$$

If magnetization and gravitomagnetization results from the application of a gravitomagnetic field to the body the relation between the magnetic and gravitomagnetic susceptibility is:

$$\frac{c_g}{c} = \frac{m_{0g}}{m_0}\left(\frac{m}{e}\right)^2 \qquad (11)$$

1. The magnetization and gravitomagnetization one obtains through the magnetic and gravitomagnetic Barnett effect (for non-ferromagnetic materials) are respectively:

$$\vec{M} = \frac{c}{m_0}\vec{B}_{equi} = -\frac{c}{m_0}\frac{2}{g}\frac{m}{e}\vec{\Omega} \qquad (12)$$

$$\vec{M}_g = \frac{c_g}{m_{0g}}\vec{B}_{g\,equi} = -\frac{c_g}{m_{0g}}\frac{2}{g}\vec{\Omega} \qquad (13)$$

The magnetization and gravitomagnetization one obtains through the rotation of a (non-ferromagnetic) body by aligning mechanically the angular momentum of the gyrostats ($e = \vec{\ell}\cdot\vec{\Omega}$) is similar to the magnetization and the gravitomagnetization one obtains by applying an external magnetic field $\vec{B}_{equi}$ (which will align the magnetic moments of the gyrostates, $e = \vec{m}\cdot\vec{B}_{equi}$) or by applying an external gravitomagnetic field $\vec{B}_{g\,equi}$ (which will align the gravitomagnetic moments of the gyrostats, $e = \vec{m}_{gm}\cdot\vec{B}_{g\,equi}$) on the same body at rest (with no rotation).

2. Let us suppose the body is at rest and non-rotating. We apply a magnetic field $B_{equi} = -\frac{1}{g}\frac{2m}{e}\Omega$ on it. What will be the magnetization and gravitomagnetization resulting from this process?

$$\vec{M} = -\frac{c}{m_0}\frac{2}{g}\frac{m}{e}\vec{\Omega} \qquad (14)$$

Having in attention Equ (9) and (10) we can compute the associated gravitomagnetic vector

$$\vec{M}_g = -\frac{c}{m_0}\frac{2}{g}\left(\frac{m}{e}\right)^2 \vec{\Omega} \qquad (15)$$

Using Equ (7) we see that this gravitomagnetization vector will be associated with a gravitomagnetic field:

$$\vec{B}_g = \frac{m_{0g}}{m_0}\frac{2}{g}\left(\frac{m}{e}\right)^2 \Omega \qquad (16)$$

which is much weaker than $\vec{B}_{g\,equi}$.

3. Let us now consider the same body at rest and non-rotating. We apply an external gravitomagnetic field of intensity $\vec{B}_{g\,equi}$. Let us compute the resulting magnetization and gravitomagnetization

$$\vec{M}_g = -\frac{c_g}{m_{0g}}\frac{2}{g}\vec{\Omega} \qquad (17)$$

Having in attention Equ (11) we can write Equ (16) under the following form:

$$\vec{M}_g = -\frac{c}{m_0}\frac{2}{g}\left(\frac{m}{e}\right)^2 \vec{\Omega} \qquad (18)$$

now applying the inverse of Equ (9) to (17) we get

$$\vec{M} = -\frac{c}{m_0}\frac{2}{g}\frac{m}{e}\vec{\Omega} \qquad (19)$$

Using the inverse of Equ (7) we can see that this magnetization vector will be associated with a magnetic field:

$$\vec{B}_g = \frac{m_0}{m_{0g}} \frac{2}{g} \frac{e}{m} \Omega \qquad (20)$$

which is much higher than $\vec{B}_{equi}$.

Comparing the magnetization and gravitomagnetization processes in point 1, 2 and 3 we see that we always end with the same magnetization and gravitomagnetization vectors despite the fact that the associated couple of magnetic and gravitomagnetic fields have not at all the same value. Therefore the three different processes are equivalent from the point of view of magnetization and gravitomagnetization but are not at all equivalent from the point of view of the intensity of the magnetic and gravitomagnetic fields involved.

Therefore the gravitomagnetic Barnett effect would be the only physical effect which would allow the production of important gravitomagnetic fields despite the fact of being associated with bodies having an extremely small gravitomagnetization.

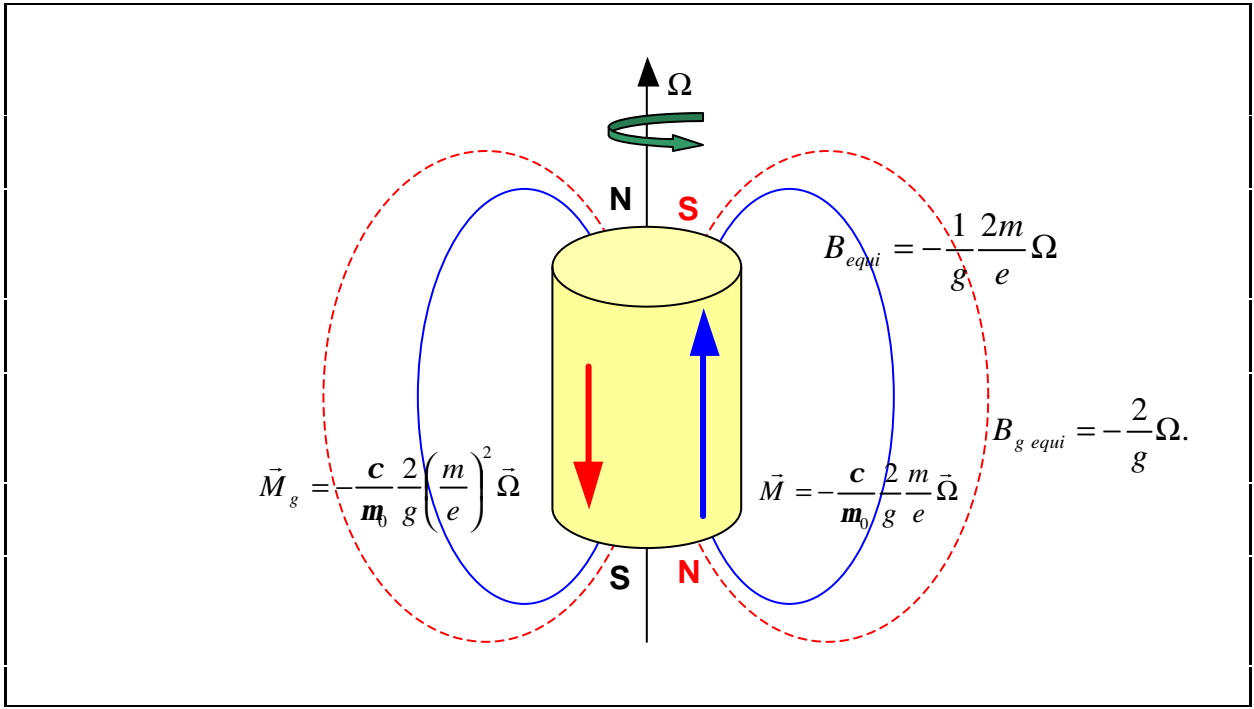

**Figure 1:** A rotating non-ferromagnetic cylinder with its associated magnetic and gravitomagnetic equivalente fields and their respective magnetization and gravitomagnetization vectors

## V - Conclusion

As stated above in section II for the magnetic Barnett effect, we may say that a gyrogravitomagnetic experiment gives no information at all concerning the process of gravitomagnetization of the rotating substance. However the presence of the Landé factor $g$ in Equ (5) indicates that the conversion of an angular velocity into a gravitomagnetic field is operated through a quantum process.

The value of the Landé factor in the context of gravitomagnetism, and the possible processes of gravitomagnetization[2,8] of different substances shall be evaluated. In the case of non-detection of the predicted effect, we might conclude that the concept of quantum gravitomagnetic moment does not make sense. This would impose important constraints on the theories of quantized gravity being currently elaborated.

Therefore, further theoretical and experimental investigation is required to confirm or not the predicted gravitomagnetic Barnett effect. This investigation is justified amongst other reasons by the macroscopic value of the gravitomagnetic field involved in the effect.